\documentclass[12pt]{article}

\usepackage{graphicx}
\usepackage{amsmath,amsthm,amssymb,amscd,a4wide}
\usepackage{dsfont}
\usepackage{color}

\DeclareMathOperator{\rank}{rank}

\DeclareMathOperator{\diag}{diag}
\DeclareMathOperator{\bv}{bv}

\numberwithin{equation}{section}
\theoremstyle{definition}
\newtheorem{theorem}{Theorem}[section]
\newtheorem{thm}[theorem]{Theorem}

\newtheorem{defn}[theorem]{Definition}

\newenvironment{psmallmatrix}
  {\left(\begin{smallmatrix}}
  {\end{smallmatrix}\right)}

\begin{document}

\begin{center}

{\LARGE\bf Exactly solvable interacting two-particle\\[5mm]
quantum graphs} \\

\vspace*{1cm}

{\large Jens Bolte}%
\footnote{E-mail address: {\tt jens.bolte@rhul.ac.uk}}
{\large and  George Garforth}%
\footnote{E-mail address: {\tt george.garforth.2012@live.rhul.ac.uk}}
\vspace*{1cm}

Department of Mathematics\\
Royal Holloway, University of London\\
Egham, TW20 0EX\\
United Kingdom\\

\end{center}

\vspace*{1cm}

\begin{abstract}
We construct models of exactly solvable two-particle quantum graphs with certain non-local two-particle interactions, establishing appropriate boundary conditions via suitable self-adjoint realisations of the two-particle Laplacian. Showing compatibility with the Bethe ansatz method, we calculate quantisation conditions in the form of secular equations from which the spectra can be deduced. We compare spectral statistics of some examples to well known results in random matrix theory, analysing the chaotic properties of their classical counterparts.
\end{abstract}

\section{Introduction}\label{sex:intro}

In this paper we investigate the properties of two-particle quantum graphs with particular focus on the acquisition and analysis of their spectra. A quantum graph is a collection of vertices and edges of finite or infinite length equipped with a differential operator. The first theoretical model of a quantum graph was devised by Pauling \cite{Pauling36}. His motivation was to study the dynamics of free electrons in hydrocarbons by modelling carbon molecules as vertices and carbon-carbon bonds as edges. This idea was later adopted by Ruedenberg and Scherr \cite{RS53} who used quantum graphs to describe free electrons donated by covalent bonds confined to entire quasi-one-dimensional molecules. Since then there have been multiple applications of quantum graphs in a variety of fields including quantum waveguides \cite{FJK87}, quantum chaos \cite{KS97}, quantum computation \cite{Lovett09} and mesoscopic systems \cite{TM10}. For a review of quantum graphs, see \cite{Exnetal08,BerKuc13}.

An important aspect of quantum graphs is that they may serve as models for quantum systems with corresponding complex classical dynamics. Kottos and Smilansky \cite{KS97} demonstrated that eigenvalue correlations in quantum graphs can be described with random matrix models, therefore providing an example for the celebrated Bohigas-Giannoni-Schmit conjecture \cite{BohGiaSch84} which is a central topic in quantum chaos. While the majority of quantum graphs literature is focussed on one-particle models, there have been a number of studies of many-particle quantum graphs. The first of these, by Melnikov and Pavlov \cite{MP95}, investigated the dynamics of two interacting particles on a connected graph with three infinite edges. Under certain restrictions of the system, they were able to find self-adjoint realisations of the two-body Laplacian corresponding to particle-particle and particle-vertex interactions. The resulting two-body wave function allowed the calculation of the conductivity of the system. More recently, Bolte and Kerner constructed two-particle quantum graphs with interactions localised at the vertices \cite{BKSingular} and later with singular contact interactions \cite{BKContact}. Boundary conditions via suitable self-adjoint extensions of the two-particle Laplacian were ascertained using quadratic forms. These results were then used to study Bose-Einstein condensation \cite{BolKerBEC}.

To some extent, the success of one-particle quantum graph models relies upon the fact that their spectra are determined by a secular equation \cite{KS97}, i.e., quantum eigenvalues are given as zeros of a finite-dimensional determinant. This fact leads to very efficient methods to calculate eigenvalues, and also allows one to prove exact trace formulae for spectral densities \cite{Rot83,KS97,BolEnd09}. The reason behind this is the fact that, locally, the classical configuration space of a one-particle graph is one dimensional. For the quantum model this means that every eigenfunction must be a linear combination of left- and right-moving, one dimensional plane waves. Many-particle quantum graphs have higher dimensional classical configuration spaces, in general prohibiting a finite dimensional secular equation that determines the eigenvalues. This obstacle can be overcome under specific circumstances when symmetries lead to an exactly solvable model.

The first model of an exactly solvable many-body quantum system confined to a single dimension is due to Lieb and Liniger \cite{LL63}. They determined the exact spectra of a repulsively $\delta$-interacting Bose gas on a circle, a result which was later generalised to distinguishable particles by Yang \cite{Ya67}. Gaudin \cite{Gau71} later employed the Bethe ansatz to describe similar systems confined to an interval. These methods were formalised by the use of the Bethe ansatz, a sum of two-particle plane waves over possible particle configurations. Implicit in the use of the Bethe ansatz is the requirement for certain symmetries brought about by the interactions in the model. The consequence of increasing the complexity to systems of particles on two-edge graphs is that these symmetries are destroyed. By imposing certain non-local particle interactions, however, Caudrelier and Cramp\'{e} \cite{CC07} showed that compatibility with the Bethe ansatz is recovered. They were then able to calculate the exact spectra of these two-edge many-particle quantum graphs. Extending this method to general two-particle quantum graphs is the main aim of this paper.

In the next section we review the construction of general one-particle quantum graphs establishing appropriate boundary conditions by self-adjoint realisation of the one-particle Laplacian $-\Delta_1$. The boundary conditions lead to a quantisation condition from which the spectra can be calculated. In Section \ref{sec:2PQG}, we review the construction of two-particle quantum graphs introducing the notion of $\delta$-interactions in terms of boundary conditions which characterise a self-adjoint Laplacian $-\Delta_2$. We introduce the Bethe ansatz method for systems of two particles on a single edge, recovering historical results mentioned above. Systems of $\delta$-interacting particles on graphs with more than a single edge are, in general, incompatible with the Bethe ansatz method. In Section \ref{sec:2PES}, we construct exactly solvable systems of two particles on equilateral stars introducing the notion of $\tilde{\delta}$-type interactions. These will be generalisations of the interactions imposed in \cite{CC07} to ensure exact solvability in the two-edge setting. Establishing appropriate boundary conditions of the two-particle Laplacian, we prove exact solvability and calculate the spectra using the Bethe ansatz. In Section \ref{sec:2PESQG} we extend our argument from equilateral stars to general graphs. In Section \ref{sec:SS} we analyse the spectral statistics of a number of examples of two-particle quantum graphs and comment on the nature of the corresponding classical dynamics.

\section{One-particle quantum graphs}\label{sec:1PQG}

In this section we review one-particle quantum graphs and their spectral properties. A combinatorial, oriented graph $\Gamma(\mathcal{V},\mathcal{I},\mathcal{E},f)$ is a collection of vertices $\mathcal{V}=\{v_1,\dots,v_{|\mathcal{V}|}\}$, connected by a set of internal edges $\mathcal{I}=\{i_1,\dots,i_{|\mathcal{I}|}\}$ and external edges $\mathcal{E}=\{e_1,\dots,e_{|\mathcal{E}|}\}$. The map $f$ assigns to each external edge $e_j$ a single vertex $f(e_j)=v_\eta$, and to each internal edge $i_j$ an ordered pair of vertices $f(i_j)=(v_\gamma,v_\lambda)$ where $v_\gamma=:f_0(i_j)$ and $v_\lambda=:f_l(i_j)$ are initial and terminal vertices respectively. A pair of edges will be called distant if they have no common vertex and neighbouring if they have at least one common vertex. The set of distant and neighbouring edge couples will be denoted $\mathcal{D}$ and $\mathcal{N}$, respectively. The degree $d_\eta$ of a vertex $v_\eta\in\mathcal{V}$ is the number of edges connected to it. The combinatorial graph is turned into a metric graph by assigning a finite interval $[0,l_j]$ to each internal edge $i_j\in\mathcal{I}$ in such a way that $f_0(i_j)$ is identified with $x=0$ and $f_l(i_j)$ with $x=l_j$. To each external edge $e_j\in\mathcal{E}$, a half-line $[0,\infty)$ is assigned such that $f(e_j)$ is identified with $x=0$. A metric graph is called compact if there are no external edges, $\mathcal{E}=\emptyset$.

Let us consider the metric graph associated with $\Gamma=\Gamma(\mathcal{V},\mathcal{I},\mathcal{E},f)$. The appropriate Hilbert space
\begin{align}\label{eq:OneParticleHilbert}
\mathcal{H}_1=\left(\bigoplus_{j=1}^{|\mathcal{I}|} L^2(0,l_j)\right)\oplus\left(\bigoplus_{j=|\mathcal{I}|+1}^{|\mathcal{I}|+|\mathcal{E}|}L^2(0,\infty)\right)
\end{align}
is the direct sum of the constituent Hilbert spaces on each edge. Vectors $\Psi=(\psi_j)_{j=1}^{|\mathcal{I}|+|\mathcal{E}|}\in\mathcal{H}_1$ are lists of square-integrable functions $\psi_j:(0,l_j)\to\mathbb{C}$, $j\in\{1,\dots,|\mathcal{I}|\}$ and $\psi_i:(0,\infty)\to\mathbb{C}$, $j\in\{|\mathcal{I}|+1,\dots,|\mathcal{I}|+|\mathcal{E}|\}$. A quantum graph is a metric graph $\Gamma$ with an associated Laplacian $-\Delta_1$ which acts according to
\begin{align}
-\Delta_1\Psi=\left(-\psi''_j(x)\right)_{j=1}^{|\mathcal{I}|+|\mathcal{E}|},
\end{align}
where dashes denote ordinary, possibly weak, derivatives. In order for the Laplacian to be a one-particle quantum Hamiltonian it needs to be realised as a self-adjoint operator.

\subsection{Self-adjoint realisation}\label{sec:SAE}
One-particle observables on $\Gamma$ are self-adjoint operators on $\mathcal{H}_1$. We thus look for self-adjoint realisations of $-\Delta_1$ with domains characterised by boundary conditions at the vertices. To this end we define boundary vectors $\Psi_{bv},\Psi_{bv}'\in\mathbb{C}^{2|\mathcal{I}|+|\mathcal{E}|}$ according to
\begin{align}\label{eq:BV}
\Psi_{\bv}=
\begin{pmatrix}
\left(\psi_j(0)\right)_{j=1}^{|\mathcal{I}|}\\
\left(\psi_j(l_j)\right)_{j=1}^{|\mathcal{I}|}\\
\left(\psi_j(0)\right)_{j=|\mathcal{I}|+1}^{|\mathcal{I}|+|\mathcal{E}|}
\end{pmatrix}
\text{ and }
\Psi'_{\text{bv}}=
\begin{pmatrix}
\left(\psi'_j(0)\right)_{j=1}^{|\mathcal{I}|}\\
\left(-\psi'_j(l_j)\right)_{j=1}^{|\mathcal{I}|}\\
\left(\psi'_j(0)\right)_{j=|\mathcal{I}|+1}^{|\mathcal{I}|+|\mathcal{E}|}.
\end{pmatrix}
\end{align}
Letting $H^2(\Gamma)$ be the set of all $\Psi\in\mathcal{H}_1$ such that $\psi_j \in H^2(0,l_j)$ for all $j\in\{1,\dots,|\mathcal{I}|\}$ and $\psi_j \in H^2(0,\infty)$ for all $j\in\{|\mathcal{I}|+1,\dots,|\mathcal{I}|+|\mathcal{E}|\}$, we can state the following theorem \cite{KosSch99}.
\begin{thm}\label{thm:1PSA}
The Laplacian $-\Delta_1$ is self-adjoint on the set of all $\Psi\in H^2(\Gamma)$ which satisfy the boundary condition
\begin{align}\label{eq:BoundaryConditions}
A\Psi_{bv}+B\Psi'_{bv}=0
\end{align}
with $(2|\mathcal{I}|+|\mathcal{E}|)\times (2|\mathcal{I}|+|\mathcal{E}|)$ matrices $A,B$ subject to the conditions $\rank(A,B)=2|\mathcal{I}|+|\mathcal{E}|$ and $AB^\dagger=BA^\dagger$.
\end{thm}

\subsection{Spectra of one-particle quantum graphs}\label{sec:1pSpectra}

The task is then to calculate the spectra of these quantum graphs by considering the eigenvalue equation
\begin{align}
-\Delta_1\Psi=E\Psi
\end{align}
alongside boundary conditions prescribed by Theorem \ref{thm:1PSA}. We shall focus on compact graphs where the spectra are discrete.

The starting point is the observation that the components $\psi_j$ of eigenfunctions 
$\Psi\in \mathcal{H}_1$ with real Laplace eigenvalues $E=k^2$ are necessarily of the form
\begin{align}\label{eq:1PAnsatz}
\psi_j(x)=\alpha_j e^{ikx}+\beta_j e^{-ikx},
\end{align}
where $\alpha_j$ and $\beta_j$ are complex constants. Imposing boundary conditions \eqref{eq:BoundaryConditions} on functions $\Psi\in H^2(\Gamma)$ and defining matrices
\begin{align}\label{eq:1PSR}
S_v(k)=-(A+ikB)^{-1}(A-ikB)
\end{align}
and
\begin{align}\label{eq:Tl}
T(k,\boldsymbol{l})=
\begin{pmatrix}
0&e^{ik\boldsymbol{l}}\\
e^{ik\boldsymbol{l}}&0
\end{pmatrix}
\end{align}
with
\begin{align}\label{eq:Boldl}
e^{ik\boldsymbol{l}}=\diag(e^{ikl_j})_{j=1}^{|\mathcal{I}|},
\end{align}
we can state the following theorem \cite{KosSch05}.
\begin{thm}\label{thm:Spectra}
The non-zero eigenvalues of a self-adjoint Laplacian $-\Delta_1$ defined on $\Gamma$ and specified through $A,B$ are the values $E=k^2$ with multiplicity $m$, where $k\neq 0$ are solutions to the secular equation
\begin{align}\label{eq:1PSecular}
\det\left[\mathbb{I}_{2|\mathcal{I}|}-S_v(k)T(k,\boldsymbol{l})\right]=0
\end{align}
with multiplicity $m$.
\end{thm}

\subsection{Star representation}\label{sec:1pScatteringMatrices}

We have seen that the spectrum of a compact quantum graph is given by the secular equation \eqref{eq:1PSecular} which is a function of matrices $T(k,\boldsymbol{l})$ and $S_v(k)$. The former clearly contains the metric information. The latter contains information about the interactions at the vertices prescribed by $A$ and $B$. In what follows we restrict our attention to \textit{local} boundary conditions where boundary values of functions at different vertices are not related. The significance of this is that we can consider scattering at each vertex independently. We formalise this interpretation by dissecting the compact graph into a collection of star graphs with finitely many, infinite edges.

\begin{defn}\label{def:SGrep}
Consider a compact graph $\Gamma(\mathcal{V},\mathcal{I},f)$. Let the map $g$ associate to each internal edge $i_j$ an ordered pair of external edges $g(i_j):=(e_{j},e_{j+|\mathcal{I}|})$. Here $e_{j}=:g_0(i_j)$ and $e_{j+|\mathcal{I}|}=:g_l(i_j)$ are external edges associated with initial and terminal vertices of $i_j$ respectively so that $f(e_{j})=f_0(i_j)$ and $f(e_{j+|\mathcal{I}|})=f_l(i_j)$. The star representation of the compact graph $\Gamma$ is the collection $\Gamma^{(s)}(\mathcal{V},\mathcal{E},f)$ of star graphs $\Gamma_\eta(v_\eta,\mathcal{E}_\eta,f)$ where $\mathcal{E}_\eta$ is the set of edges $e_j$ such that $f(e_j)=v_\eta$. Clearly we have that $2|\mathcal{I}|=|\mathcal{E}|$. The star graphs are turned into metric graphs by assigning half-lines $[0,\infty)$ to its edges.
\end{defn}

Consider the star representation $\Gamma^{(s)}$ of a compact graph $\Gamma$. The Hilbert space associated with $\Gamma^{(s)}$ is
\begin{align}
\mathcal{H}_1^{(s)}=\bigoplus_{j=1}^{|\mathcal{E}|}L^2(0,\infty),
\end{align}
and boundary values of vectors $\Psi=(\psi_j^{(s)})_{j=1}^{|\mathcal{E}|}\in\mathcal{H}_1^{(s)}$ are
\begin{align}
\Psi_{\bv}^{(s)}=\left(\psi_j^{(s)}(0)\right)_{j=1}^{|\mathcal{E}|}\quad\text{and}\quad
{\Psi_{\bv}^{(s)}}'=\left({\psi_j^{(s)}}'(0)\right)_{j=1}^{|\mathcal{E}|},
\end{align}
so that analogues of boundary conditions \eqref{eq:BoundaryConditions} are given by
\begin{align}\label{eq:LocalBoundaryConditions}
A\Psi_{\bv}^{(s)}+B{\Psi_{\bv}^{(s)}}'=0.
\end{align}
Let  $\mathbb{P}$ be an $|\mathcal{E}|$-dimensional permutation matrix which reorders vectors $\Psi$ according to
\begin{align}\label{eq:P}
\mathbb{P}\Psi=(\Psi_{\eta})_{\eta=1}^{|\mathcal{V}|},
\end{align}
where each $\Psi_{\eta}$ lists functions $\psi^{(s)}_j$ with $f(e_j)=v_\eta$. Local boundary conditions then imply the decomposition
\begin{align}\label{eq:LocalAB}
A=\mathbb{P}^{-1}\left(\bigoplus_{v_\eta\in \mathcal{V}}A_\eta\right)\mathbb{P} \text{ and } B=\mathbb{P}^{-1}\left(\bigoplus_{v_\eta\in \mathcal{V}}B_\eta\right)\mathbb{P}.
\end{align}
The total scattering matrix can be reconstructed from sub-graphs $\Gamma_\eta$ according to
\begin{align}
S_v(k)=\mathbb{P}^{-1}\left(\bigoplus_{\eta=1}^{|\mathcal{V}|}S_v^{(\eta)}(k)\right)\mathbb{P}
\end{align}
with
\begin{align}
S_v^{(\eta)}(k)=-(A_\eta+ikB_\eta)^{-1}(A_\eta-ikB_\eta).
\end{align}

Let us retrieve the secular equation \eqref{eq:1PSecular} by reconstructing the original compact graph from its star representation (see \cite{KS97,KurNow05}). Consider the functions $\psi^{(s)}_j$ and $\psi^{(s)}_{j+|\mathcal{I}|}$ related to the external edges $e_j$ and $e_{j+|\mathcal{I}|}$ respectively. Joining up the external edges to form a single internal edge of length $l_j$ is imposing the relation
\begin{align}
\psi^{(s)}_j(x)=\psi^{(s)}_{j+|\mathcal{I}|}(l_j-x).
\end{align}
Choosing the form \eqref{eq:1PAnsatz} and defining vectors
\begin{align}
\boldsymbol{\alpha}=(\alpha_j)_{j=1}^{|\mathcal{E}|}\text{ and }\boldsymbol{\beta}=(\beta_j)_{j=1}^{|\mathcal{E}|}
\end{align}
we have the relation
\begin{align}\label{eq:1particlestarp}
\boldsymbol{\beta}=T(k,\boldsymbol{l})\boldsymbol{\alpha}.
\end{align}
The total scattering matrix in this context acts according to
\begin{align}\label{eq:1particlestarv}
\boldsymbol{\alpha}=S_v(k)\boldsymbol{\beta}.
\end{align}
Applying \eqref{eq:1particlestarp} and \eqref{eq:1particlestarv} successively we recover the secular equation \eqref{eq:1PSecular} as required.

\section{Two-particle quantum graphs with $\delta$-type interactions}\label{sec:2PQG}

The Hilbert space of a many-particle quantum system is given by the tensor product of one-particle Hilbert spaces. The appropriate two-particle Hilbert space for a compact two-particle quantum graph is
\begin{align}
\mathcal{H}_2=\left(\bigoplus_{j=1}^{|\mathcal{I}|}L^2(0,l_j)\right)\otimes\left(\bigoplus_{j=1}^{|\mathcal{I}|}L^2(0,l_j)\right)=\bigoplus_{(i_m,i_n)\in\mathcal{I}\times\mathcal{I}}L^2(D_{mn}),
\end{align}
where
\begin{align}\label{eq:DomainBK}
D_{mn}=(0,l_m)\times (0,l_n).
\end{align}
The total classical configuration space for two particles on $\Gamma$ is the union 
\begin{align}\label{eq:TotalDomainBK}
D_\Gamma=\bigcup_{m,n=1}^{|\mathcal{I}|}D_{mn}
\end{align}
of rectangles. The two-particle Hilbert space can then be written $\mathcal{H}_2=L^2(D_\Gamma)$. Vectors $\Psi=(\psi_{mn})_{m,n=1}^{|\mathcal{I}|}$ consist of functions $\psi_{mn}:D_{mn}\to\mathbb{C}$. 

At this point let us introduce the two-particle Laplacian $-\Delta_2$ which acts according to
\begin{align}
-\Delta_2\Psi=\left(-\frac{\partial^2\psi_{mn}}{\partial{x_1^2}}-\frac{\partial^2\psi_{mn}}{\partial{x_2^2}}\right)_{m,n=1}^{|\mathcal{I}|}.
\end{align}
We wish to consider the two-particle eigenvalue equation
\begin{align}\label{eq:2pGSE}
-\Delta_2\Psi=E\Psi
\end{align}
alongside boundary conditions which prescribe single-particle interactions with vertices as well as singular contact interactions between particles. The latter interactions take place along the diagonals $x_1=x_2$ of squares $D_{mm}$ and are rigorous versions of $\delta$-interactions of the form
\begin{align}\label{deltainteract}
\alpha\delta\left(x_1 -x_2\right).
\end{align}
They naturally define the dissected configuration space
\begin{align}D_\Gamma^*=\left(\bigcup_{m,n=1|m\neq n}^{\mathcal{|I|}} D_{mn}\right)\bigcup \left(\bigcup_{m=1}^{\mathcal{|I|}} \left(D_{mm}^+\cup D_{mm}^-\right)\right) 
\end{align}
with subdomains of squares $D_{mm}$ defined
\begin{align}
D_{mm}^+=\{(x_1,x_2)\in D_{mm};x_1>x_2\}
\end{align}
and
\begin{align}
D_{mm}^-=\{(x_1,x_2)\in D_{mm};x_1<x_2\}.
\end{align}
The total dissected two-particle Hilbert space is then $\mathcal{H}_2^*=L^2(D_\Gamma^*)$. Thus two-particle wave functions $\Psi\in\mathcal{H}_2^*$ are lists
\begin{align}\label{eq:2particlevector}
\Psi=
\begin{pmatrix}
(\psi_{mn})_{m,n=1|m\neq n}^{|\mathcal{I}|}\\
(\psi_{mm}^+)_{m=1}^{|\mathcal{I}|}\\
(\psi_{mm}^-)_{m=1}^{|\mathcal{I}|}
\end{pmatrix}
\end{align}
of functions $\psi_{mn}:D_{mn}\to\mathbb{C}$, $m\neq n$, and $\psi^\pm_{mm}:D^\pm_{mm}\to\mathbb{C}$. The space $H^2 (D_\Gamma^*)$ is defined in an analogous way.

The boundary conditions that one needs to impose on vectors $\Psi\in H^2(D_\Gamma^*)$ in order to generate single-particle interactions with the vertices as well as singular contact interactions \eqref{deltainteract} were established in \cite{BKContact} and consist of two parts. The first set of boundary conditions is as described in Theorem \ref{thm:1PSA} for each of the two variables separately. The boundary values of $\Psi\in H^2 (D_\Gamma^*)$ at the vertices are
\begin{align}\label{eq:ContactVVector}
\Psi_{\bv}^{(v)}(y)=
\begin{pmatrix}
\left(\psi_{mn}(0,l_ny)\right)_{m,n=1}^{|\mathcal{I}|}\\
\left(\psi_{mn}(l_m,l_ny)\right)_{m,n=1}^{|\mathcal{I}|}\\
\left(\psi_{mn}(l_my,0)\right)_{n,m=1}^{|\mathcal{I}|}\\
\left(\psi_{mn}(l_my,l_n)\right)_{n,m=1}^{|\mathcal{I}|}
\end{pmatrix}
\text{ and }
{\Psi_{\bv}^{(v)}}'(y)=
\begin{pmatrix}
\left(\psi_{mn,1}(0,l_ny)\right)_{m,n=1}^{|\mathcal{I}|}\\
\left(\psi_{mn,1}(l_m,l_ny)\right)_{m,n=1}^{|\mathcal{I}|}\\
\left(\psi_{mn,2}(l_my,0)\right)_{n,m=1}^{|\mathcal{I}|}\\
\left(\psi_{mn,2}(l_my,l_n)\right)_{n,m=1}^{|\mathcal{I}|}
\end{pmatrix}
\end{align}
for all $y\in(0,1)$ where, for compactness the labels $\pm$ are dropped. Boundary conditions at the vertices are then
\begin{align}\label{eq:DeltaV}
\left(\mathbb{I}_2\otimes A\otimes \mathbb{I}_{|\mathcal{I}|}\right)\Psi_{\bv}^{(v)}+\left(\mathbb{I}_2\otimes B\otimes \mathbb{I}_{|\mathcal{I}|}\right){\Psi_{\bv}^{(v)}}'=0.
\end{align}
The $\delta$-type interaction conditions are
\begin{align}\label{eq:DeltaP1}
\psi_{mm}^+(x_1,x_2)_{x_1=x_2^+}&=\psi_{mm}^-(x_1,x_2)_{x_1=x_2^-};\\
\label{eq:DeltaP2}\left(\frac{\partial}{\partial x_1}-\frac{\partial}{\partial x_2}-2\alpha\right)\psi_{mm}^+(x_1,x_2)|_{x_1=x_2^+}&=\left(\frac{\partial}{\partial x_1}-\frac{\partial}{\partial x_2}\right)\psi_{mm}^-(x_1,x_2)|_{x_1=x_2^-}
\end{align}
for every $x\in(0,l_m)$. Here, and in the remainder of this paper, we use the notation $x^\pm=\lim_{\delta\rightarrow 0}(x\pm\delta)$. In \cite{BKContact} it was proven that the two-particle Laplacian with these boundary conditions is self-adjoint.

In Section \ref{sec:1PQG}, we calculated the spectra of one particle quantum graphs by specifying the form \eqref{eq:1PAnsatz} of eigenfunctions of $-\Delta_1$ and applying boundary conditions \eqref{eq:BoundaryConditions}. We would like to extend this approach to the two-particle quantum graph setting. As we are dealing with a two dimensional configuration space, the difficulty is that in general there does not exist a suitable analogue of the general form of an eigenfunction \eqref{eq:1PAnsatz}. It is well known, however, that in particular cases a Bethe ansatz can be used in this way.

Let us begin by considering systems of $\delta$-interacting particles on the simplest metric graph, an interval $[0,l]$. Such systems were exactly solved in \cite{Gau71,Ya67,LL63,Bet31} using the Bethe ansatz; a superposition of possible many-particle plane wave states as explicit eigenfunctions of the two-particle Laplacian. In this way exact spectra of many-particle systems can be calculated. To see how this works in the present setting we choose Dirichlet boundary conditions at the interval ends, $A=\mathbb{I}_2$ and $B=0$. The Bethe ansatz method in this context is the assumption that eigenfunctions of $-\Delta_2$ take the form
\begin{align}\label{eq:BoxAnsatz}
\psi^\pm(x_1,x_2)=\sum_{P\in\mathcal{W}_2}\mathcal{A}^{(P,\pm)} e^{i(k_{P1} x_1+k_{P2} x_2)},
\end{align} 
with amplitudes $\mathcal{A}^{(P,\pm)}$ and where elements $P$ of the Weyl group
\begin{align}
\mathcal{W}_2=(\mathbb{Z}/2\mathbb{Z})^2\rtimes S_2
\end{align}
act according to $k_{Pi}=\sigma_P k_{Qi}$, with $\sigma_P=\pm 1$ and $Q\in S_2$ for $i\in\{1,2\}$. Each of the eight elements $P$ of the Weyl group $\mathcal{W}_2$ can be written in terms of the two generators $R$ and $T$ which satisfy the conditions
\begin{enumerate}
\item $TT=I$;
\item $RR=I$;
\item $TRTR=RTRT$.
\end{enumerate}
The eigenvalue equation \eqref{eq:2pGSE} is then satisfied with Laplace eigenvalues $k_1^2+k_2^2$.

We would first like to verify that the system is indeed exactly solvable, that is, boundary conditions \eqref{eq:DeltaV}--\eqref{eq:DeltaP2} imposed on the ansatz \eqref{eq:BoxAnsatz} are compatible with these conditions. To this end let us define the $2$-dimensional vector
\begin{align}
\mathcal{A}^P=\begin{pmatrix}
\mathcal{A}^{(P,-)}\\
\mathcal{A}^{(PT,+)}
\end{pmatrix}.
\end{align}
The vertex boundary conditions \eqref{eq:DeltaV} then imply
\begin{align}\label{eq:PR}
\mathcal{A}^{PR}=-\mathcal{A}^P \text{ and }\mathcal{A}^{PTRT}=-e^{2ik_{P2}l}\mathcal{A}^P
\end{align}
for all $P\in\mathcal{W}_2$. The $\delta$-type interaction conditions \eqref{eq:DeltaP1} and \eqref{eq:DeltaP2} imply
\begin{align}\label{eq:PT}
\mathcal{A}^{PT}=S_p(k_{P1}-k_{P2})\mathcal{A}^P
\end{align}
with
\begin{align}\label{eq:Sp}
S_p(k)=\frac{1}{k+i\alpha}\begin{pmatrix}
-i\alpha & k\\
k & -i\alpha
\end{pmatrix}.
\end{align}
To prove exact solvability we need only show that relations \eqref{eq:PR} and \eqref{eq:PT} are consistent with the properties of $\mathcal{W}_2$. This amounts to the requirements
\begin{enumerate}
\item $S_p(u)S_p(-u)=\mathbb{I}_2$,
\item $S_p(u)S_p(v)=S_p(v)S_p(u)$,
\end{enumerate}
which are easily verified by the explicit form of $S_p(k)$. Now we have established that the system is exactly solvable, we would like to deduce the spectrum. Applying \eqref{eq:PR} and \eqref{eq:PT} successively, we arrive at the quantisation condition that
\begin{align}\label{eq:Sec}
Z_\text{interval}(k_{P1},k_{P2})=0
\end{align}
with
\begin{align}\label{eq:QCForm}
Z_\text{interval}(k_1,k_2)=\det\left[\mathbb{I}_{2}-e^{2ik_{1}l}S_p(k_{1}-k_{2})S_p(k_{1}+k_{2})\right]
\end{align}
is satisfied for all $P\in\mathcal{W}_2$. Noting here that the form of $S_p(k)$ is such that if \eqref{eq:Sec} is satisfied for some $P\in\mathcal{W}_2$, then it is necessarily satisfied for elements $PR,PTRT\in\mathcal{W}_2$, we have the following result first established in \cite{Gau71}.
\begin{thm}\label{thm:BoxSpectrum}
Eigenvalues of a self-adjoint two-particle Laplacian $-\Delta_2$ defined on an interval $[0,l]$ with Dirichlet interactions at the endpoints and $\delta$-type particle interactions are the values $E=k_1^2+k_2^2 \neq 0$ with multiplicity $m$ where $k_1,k_2$ are simultaneous solutions to the secular equations
\begin{align}\label{eq:BoxFinalQCon}
Z_{\text{interval}}(k_i,k_j)=0
\end{align}
for $j,i\neq j\in\{1,2\}$ with multiplicity $m$.
\end{thm}
We note that, instead choosing boundary conditions
\begin{align}
A=\begin{pmatrix}
1&-1\\
0&0
\end{pmatrix}
\text{ and }
B=\begin{pmatrix}
0&0\\
1&1
\end{pmatrix}
\end{align}
we recover the quantisation condition for two particles on a circle deduced in \cite{Ya67}. Furthermore, imposing bosonic symmetry
\begin{align}\label{eq:BosSym}
\psi_{mm}^\pm(x_1,x_2)=\psi_{mm}^\mp(x_1,x_2)
\end{align}
we recover the result in \cite{LL63}.

The aim of this paper is to calculate spectra for general two-particle graphs. It turns out that systems of $\delta$-interacting particles on graphs with more than a single edge, in general, are not exactly solvable in this way. The task is then to establish boundary conditions on general quantum graphs which are compatible with the Bethe ansatz method. In \cite{CC07}, systems of particles on two-edge stars, interacting via certain non-local $\delta$-type interactions, were shown to be exactly solvable. Then prescribing a length $l$ to the edges and imposing certain coupling conditions, the spectra were deduced. In what follows, we extend this approach to systems of two particles on general graphs. Furthermore, we show that the corresponding boundary conditions are compatible with a self-adjoint Laplacian.

\section{Exactly solvable two-particle equilateral stars}\label{sec:2PES}

Before discussing general graphs it is convenient to consider a subset of graphs called equilateral stars. These graphs exhibit most of the essential features of the general case and thus act as a convenient way to introduce some key concepts. Let us define the equilateral star $\Gamma_e$ as the graph $\Gamma(\mathcal{V},\mathcal{I},f)$ with the restrictions $l_j=l$,$f_0(i_j)=v_1$ and $f_l(i_j)=v_{j+1}$ for all $i_j\in\mathcal{I}$. The degree $d$ of the central vertex $v_1$ then is the same as the number of edges. Vectors $\Psi\in\mathcal{H}_2$ are lists of two-particle functions $\psi_{mn}:D_{mn}\rightarrow\mathbb{C}$ in $L^2(D_{mn})$ with square subdomains defined
\begin{align}
D_{mn}=(0,l)\times (0,l).
\end{align}
The total configuration space for two particles on $\Gamma_e$ is the union $D_{\Gamma_e}$ \eqref{eq:TotalDomainBK}, and the two-particle Hilbert space is $\mathcal{H}_2=L^2(D_{\Gamma_e})$. Boundary conditions at the vertices are chosen in the same way as in Section \ref{sec:2PQG}. Interactions between particles will be analogues of the non-local $\delta$-interactions imposed in \cite{CC07}. In what follow we refer to these as $\tilde\delta$-interactions. They are characterised by the conditions
\begin{align}\label{eq:TildeDelta1}
\psi_{mn}^+(x_1,x_2)|_{x_1=x_2^+}&=\psi_{nm}^-(x_1,x_2)|_{x_1=x_2^-};\\
\label{eq:TildeDelta2}\left(\frac{\partial}{\partial x_1}-\frac{\partial}{\partial x_2}-2\alpha\right)\psi_{mn}^+(x_1,x_2)|_{x_1=x_2^+}&=\left(\frac{\partial}{\partial x_1}-\frac{\partial}{\partial x_2}\right)\psi_{nm}^-(x_1,x_2)|_{x_1=x_2^-}
\end{align}
for almost every $x_i\in(0,l)$.

Using the method devised in \cite{BKContact} one can introduce vectors of boundary values that reproduce the boundary conditions \eqref{eq:TildeDelta1}--\eqref{eq:TildeDelta2} and prove that they provide a self-adjoint realisation of the two-particle Laplacian.

We stress here that $\tilde\delta$-interactions can take place when particles are located on different edges and therefore represent rather less physical interactions than $\delta$-interactions. We choose these models as they permit exact solutions via the Bethe ansatz
\begin{align}\label{eq:EquiAnsatz}
\psi_{mn}^\pm(x_1,x_2)=\sum_{P\in\mathcal{W}_2}\mathcal{A}_{mn}^{(P,\pm)} e^{i(k_{P1} x_1+k_{P2} x_2)}.
\end{align}
This form obviously leads to eigenfunctions of \eqref{eq:2pGSE} with eigenvalues $E=k_1^2+k_2^2$.

To show that the boundary conditions imposed on the ansatz \eqref{eq:EquiAnsatz} are compatible with the properties of $\mathcal{W}_2$, let us first define the $d^2\times d^2$ permutation matrix
\begin{align}\label{eq:Tmatrixdef}
\mathbb{T}_{d^2}=
\begin{pmatrix}
\mathbb{I}_d \otimes M_1\\
\vdots\\
\mathbb{I}_d \otimes M_d\\
\end{pmatrix}
\end{align}
with row vectors
\begin{align}
M_j=
\begin{pmatrix}
\underbrace{0\dots 0}_{j-1}&1&\underbrace{0\dots 0}_{d-j}
\end{pmatrix}.
\end{align}
It is convenient to note the properties 
\begin{align}
\mathbb{T}_{d^2}(\mathcal{A}^P_{mn})_{m,n=1}^{d}=(\mathcal{A}^P_{mn})_{n,m=1}^{d}
\end{align}
and 
\begin{align}
\mathbb{T}_{d^2}(M\otimes N)\mathbb{T}_{d^2}=N\otimes M
\end{align}
for any $d\times d$ matrices $M$ and $N$. Then, defining the $2d^2$-dimensional vector
\begin{align}
\mathcal{A}^P=\begin{pmatrix}
(\mathcal{A}_{mn}^{(P,-)})_{m,n=1}^{d}\\
\mathbb{T}_{d^2}(\mathcal{A}_{mn}^{(PT,+)})_{m,n=1}^{d}
\end{pmatrix},
\end{align}
the boundary conditions \eqref{eq:DeltaV} imply
\begin{align}\label{eq:GVint}
\begin{split}
&\left(\mathbb{I}_2\otimes A\otimes\mathbb{I}_{d}\right)\mathbb{Q}\begin{pmatrix}
\mathcal{A}^P+\mathcal{A}^{PR}\\
\mathcal{A}^{PT}e^{ik_{P1}l}+\mathcal{A}^{PRT}e^{-ik_{P1}l}
\end{pmatrix}\\
+ik_{P1}&\left(\mathbb{I}_2\otimes B\otimes\mathbb{I}_{d}\right)\mathbb{Q}
\begin{pmatrix}
\mathcal{A}^P-\mathcal{A}^{PR}\\
-\mathcal{A}^{PT}e^{ik_{P1}l}+\mathcal{A}^{PRT}e^{-ik_{P1}l}
\end{pmatrix}=0
\end{split}
\end{align}
for all $P\in\mathcal{W}_2$ where
\begin{align}
\mathbb{Q}=\begin{pmatrix}
\mathbb{I}_{d^2}&0&0&0\\
0&0&0&\mathbb{T}_{d^2}\\
0&\mathbb{I}_{d^2}&0&0\\
0&0&\mathbb{T}_{d^2}&0
\end{pmatrix}.
\end{align}
Equilateral stars have Dirichlet conditions at external vertices $v_j$, $j\geq 2$. We thus have the decomposition
\begin{align}\label{eq:BlockFormEqui}
A=
\begin{pmatrix}
A_1 & 0 \\ 
0 & A_2
\end{pmatrix}
\text{ and }
B=
\begin{pmatrix}
B_1 & 0 \\
0 & B_2
\end{pmatrix}
\end{align}
with
\begin{align}\label{eq:StarA2}
A_2=\mathbb{I}_{d}\text{ and }B_2=0
\end{align}
By using the properties of $\mathbb{T}_{d^2}$, we then have that
\begin{align}
\mathbb{Q}^{-1}\left(\mathbb{I}_2\otimes S_v(k)\otimes \mathbb{I}_{d}\right)\mathbb{Q}=\begin{pmatrix}\mathbb{I}_2\otimes S_v^{(1)}(k)\otimes\mathbb{I}_{d}&0\\
0&-\mathbb{I}_{2d^2}
\end{pmatrix}.
\end{align}
Rearranging \eqref{eq:GVint}, we can then extract the relation
\begin{align}\label{eq:EquiBvv}
\mathcal{A}^{PR}=\left(\mathbb{I}_2\otimes S_v^{(1)}(-k_{P1})\otimes \mathbb{I}_{d}\right)\mathcal{A}^P.
\end{align}
The $\tilde\delta$-conditions \eqref{eq:TildeDelta1}--\eqref{eq:TildeDelta2} imply
\begin{align}\label{eq:EquiBvp}
\mathcal{A}^{PT}=\left(S_p(k_{P1}-k_{P2})\otimes\mathbb{I}_{d^2}\right)\mathcal{A}^P
\end{align}
with $S_p(k)$ defined in \eqref{eq:Sp}. To prove exact solvability we need only show that relations \eqref{eq:EquiBvv}--\eqref{eq:EquiBvp} are consistent with the properties of $\mathcal{W}_2$. This amounts to the requirements
\begin{enumerate}
\item $S_v^{(1)}(u)S_v^{(1)}(-u)=\mathbb{I}_{d}$;
\item $S_p(u)S_p(-u)=\mathbb{I}_2$;
\item $\left(\mathbb{I}_2\otimes S_v^{(1)}(u)\otimes \mathbb{I}_{d} \right)\left( S_p(u+v)\otimes\mathbb{I}_{d^2}\right)\left(\mathbb{I}_2\otimes S_v^{(1)}(v)\otimes\mathbb{I}_{d}\right)\left(S_p(v-u)\otimes\mathbb{I}_{d^2}\right)\\
=\left(S_p(v-u)\otimes\mathbb{I}_{d^2}\right)\left(\mathbb{I}_2\otimes S_v^{(1)}(v)\otimes \mathbb{I}_{d} \right)\left( S_p(u+v)\otimes\mathbb{I}_{d^2}\right)\left(\mathbb{I}_2\otimes S_v^{(1)}(u)\otimes\mathbb{I}_{d}\right)$.
\end{enumerate}
The first two conditions are easily verified by the explicit forms of $S_v^{(1)}(u)$ and $S_p(u)$. The third follows from the result in \cite{KosSch05} that, for any $A,B$ and $u,v$, we have the commutation relation 
\begin{align}\label{eq:EquiKSCommutation}
[S_v(u),S_v(v)]=0.
\end{align}

Now we have established that the system is exactly solvable, we would like to deduce the spectrum. This can be done in a number of ways. The method we choose here generalises that used for the one-particle case in \cite{KosSch05} which we presented in Section \ref{sec:1pSpectra}. Substituting \eqref{eq:EquiBvp} into \eqref{eq:GVint} we have that
\begin{align}\label{eq:IntAXBY}
\left(\left(\mathbb{I}_2\otimes A\otimes\mathbb{I}_{d}\right)\mathbb{Q}X(k_{P1},k_{P2},l)
+ik_{P1}\left(\mathbb{I}_2\otimes B\otimes\mathbb{I}_{d}\right)\mathbb{Q}Y(k_{P1},k_{P2},l)\right)
\begin{pmatrix}
\mathcal{A}^P\\
\mathcal{A}^{PR}
\end{pmatrix}=0
\end{align}
with
\begin{align}
X(k_{1},k_{2},l)=
\begin{pmatrix}
\mathbb{I}_{2} & \mathbb{I}_{2}\\
S_p(k_{1}-k_{2})e^{ik_{1}l}&S_p(-k_{1}-k_{2})e^{-ik_{1}l}
\end{pmatrix}\otimes \mathbb{I}_{d^2}
\end{align}
and
\begin{align}
Y(k_{1},k_{2},l)=
\begin{pmatrix}
\mathbb{I}_{2} & -\mathbb{I}_{2}\\
-S_p(k_{1}-k_{2})e^{ik_{1}l}&S_p(-k_{1}-k_{2})e^{-ik_{1}l}
\end{pmatrix}\otimes \mathbb{I}_{d^2}.
\end{align}
Then by the properties of determinants, and since $\det(A+ikB)\neq 0$, we arrive at the condition that
\begin{align}\label{eq:EquiSecEqu}
Z_e(k_{P1},k_{P2})=0
\end{align}
with
\begin{align}\label{eq:EquiQCForm}
Z_e(k_1,k_2)=\det\left[\mathbb{I}_{2d}+e^{2ik_{1}l}\left(S_p(k_{1}-k_{2})S_p(k_{1}+k_{2})\otimes S_v^{(1)}(k_{1})\right)\right]=0
\end{align}
is satisfied for all $P\in\mathcal{W}_2$. By using properties of determinants and the explicit forms of $S_p(k)$ and $S_v(k)$, it is easy to see that the form \eqref{eq:EquiQCForm} is such that if \eqref{eq:EquiSecEqu} is satisfied for some $P\in\mathcal{W}_2$, then it is necessarily satisfied for elements $PR,PTRT\in\mathcal{W}_2$. With this in mind, we can state the main result of this section.
\begin{thm}\label{thm:EquiSpectrum}
Eigenvalues of a self-adjoint two-particle Laplacian $-\Delta_2$ defined on an equilateral star $\Gamma_e$ with interactions at the central vertex specified through $A_1,B_1$ and $\tilde{\delta}$-type particle interactions are the values $E=k_1^2+k_2^2\neq 0$ with multiplicity $m$ where $k_1,k_2$ are simultaneous solutions to the secular equations
\begin{align}\label{eq:EquiFinalQCon}
Z_e(k_i,k_j)=0
\end{align}
for $j,i\neq j\in\{1,2\}$ with multiplicity $m$.
\end{thm}

Before we move on to general graphs, let us establish agreement with some results discussed earlier in the paper.

Firstly, choosing $d=2$, we recover the spectrum of a  system of two particles on an interval with a central impurity as solved in \cite{CC07}. Furthermore, rather than choosing Dirichlet vertex conditions \eqref{eq:StarA2}, which specified the connectivity of an equilateral star, and instead choosing Kirchhoff boundary conditions
\begin{align}
A_2=
\begin{pmatrix}
1 & -1 \\ 
0 & 0
\end{pmatrix}
\text{ and }
B_2=
\begin{pmatrix}
0 & 0 \\
1 & 1
\end{pmatrix}
\end{align}
to establish continuity at the outer vertices, we recover the spectra of systems of two particles on a circle with an impurity also in \cite{CC07}.

Throughout this section we have used $\alpha$ to parameterise the strength of particle interactions. It is reasonable to expect then that by setting $\alpha=0$, one should arrive at separable quantisation conditions given by \eqref{eq:1PSecular} for one-particle quantum graphs. Indeed, by substituting $\alpha=0$ into the form \eqref{eq:EquiQCForm}, we recover
\begin{align}\label{eq:EquiQCForma0}
\det\left[\mathbb{I}_{d}+e^{2ikl}S_v^{(1)}(k)\right]=0
\end{align}
which is exactly the corresponding one-particle condition. It is important to point out however, that $\tilde{\delta}$-type interactions with $\alpha=0$ result in coupling between domains $D_{mn}$ and $D_{nm}$ and are thus clearly distinct from the truly non-interacting situation. For this reason we refer to such systems as \textit{pseudo-non-interacting}. The fact the one-particle condition \eqref{eq:EquiQCForma0} is recovered in this case is a result of the specific geometry of  the equilateral star. We will see in the subsequent section that this agreement does not hold for general graphs. We revisit this point in the final section of the chapter when discussing spectral statistics.

\section{Exactly solvable two-particle quantum graphs}\label{sec:2PESQG}

We have seen, in the previous section, how to construct exactly solvable models of two interacting particles on equilateral stars. The majority of quantum graphs literature, however, is concerned with the dynamics of single particles on graphs with, in general, different edge lengths. Indeed, in \cite{KS97}, rationally independent edge lengths are required to avoid degenerate energy levels and ensure spectral statistics following random matrix predictions. This section is concerned with extending the scope of our discussion to two particles on general compact graphs.

As imposing $\tilde{\delta}$-type interactions between particles on equilateral stars leads to exact solutions, it seems reasonable to assume that a suitable variant of such interactions in the general setting will also lead to exact solutions. However, general graphs bring added complications associated with edges of different length and distant vertices. The problem is then to choose an appropriate way to impose $\tilde{\delta}$-type interactions in the general setting which preserves compatibility with the Bethe ansatz method. To address this, let us consider a pair of particles on a general graph $\Gamma$ viewed in its star representation $\Gamma^{(s)}$. At any one time, the particles will be located on some pair of infinite stars $(\Gamma_\gamma,\Gamma_\lambda)$ with $\gamma,\lambda\in\{1,\dots,|\mathcal{V}|\}$. We impose that when particles are located on different stars ($\gamma\neq\lambda$), they will be independent of each other; there are no particle interactions. When, however, the particles are located on the same star ($\gamma=\lambda$), they will be allowed to interact. We postulate here that exact solvability is assured if these interactions are of $\tilde{\delta}$-type. As in previous sections, we would like to establish appropriate self-adjoint realisations of $-\Delta_2$ on the compact graph. Thus we must determine how these interactions translate to the compact setting. To this end consider a pair of particles on a neighbouring edge couple $(i_m,i_n)\in\mathcal{N}$ with coordinates $x_1\in[0,l_m]$ and $x_2\in[0,l_n]$ respectively. The orientations of the edges could be such that $f_0(i_m)=f_0(i_n)$, $f_0(i_m)=f_l(i_n)$, $f_l(i_m)=f_0(i_n)$ or $f_l(i_m)=f_l(i_n)$. The $\tilde{\delta}$-interactions prescribed above in the star representation become effective when either $x_1=x_2$, $x_1=l_n-x_2$, $l_m-x_1=x_2$ or $l_m-x_1=l_n-x_2$, respectively. This means that there is an interaction when the particles are located in the same position (when $m=n$), or when they are on different edges, $m\neq n$, the same distance away from the common vertex of these edges. The interaction will then be cut off at the smaller of the two edge lengths involved. The four cases can be mapped to the first one by changing coordinates according to
\begin{align}
(\tilde{x}_1,\tilde{x}_2)=\begin{cases}
(x_1,x_2)&\text{if }f_0(i_m)=f_0(i_n),\\
(x_1,l_n-x_2)&\text{if } f_0(i_m)=f_l(i_n),\\
(l_m-x_1,x_2)&\text{if } f_l(i_m)=f_0(i_n),\\
(l_m-x_1,l_n-x_2)&\text{if } f_l(i_m)=f_l(i_n).
\end{cases}
\end{align}
In these coordinates interactions take place when $\tilde x_1 =\tilde x_2$. The total dissected configuration space can be written
\begin{align}
D_{\Gamma}^*=\left(\bigcup_{(i_m,i_n)\in\mathcal{D}}D_{mn}\right)\bigcup\left(\bigcup_{(i_m,i_n)\in\mathcal{N_0}}(\tilde{D}_{mn}^+\cup \tilde{D}_{mn}^-)\right)
\end{align}
with subdomains defined as
\begin{align}
\tilde{D}_{mn}^+=\{(\tilde{x}_1,\tilde{x}_2)\in (0,l_m)\times (0,l_n);\ \tilde{x}_1>\tilde{x}_2\}
\end{align}
and
\begin{align}
\tilde{D}_{mn}^-=\{(\tilde{x}_1,\tilde{x}_2)\in (0,l_m)\times (0,l_n);\ \tilde{x}_1<\tilde{x}_2\}.
\end{align}

In order to define the quantum model, very much in analogy to \eqref{eq:2particlevector}--\eqref{eq:DeltaP2}, we need the Hilbert space $L^2(D_{\Gamma}^*)$ and a domain that is specified as a subspace of $H^2(D_{\Gamma}^*)$ in terms of boundary conditions. We denote the components of vectors $\Psi\in H^2(D_{\Gamma}^*)$ as $\psi_{mn}:D_{mn}\to\mathbb{C}$, when $(i_m,i_n)\in\mathcal{D}$ and $\psi_{mn}^\pm:D_{mn}^\pm\to\mathbb{C}$, when $(i_m,i_n)\in\mathcal{N}$. First defining vertex boundary vectors as in \eqref{eq:ContactVVector} we again have vertex conditions \eqref{eq:DeltaV}. With the definition $l_{mn}^-=\min(l_m,l_n)$ we have $\tilde{\delta}$-type boundary conditions
\begin{align}\label{eq:NonLocalDelta1}
\psi_{mn}^+(\tilde{x}_1,\tilde{x}_2)|_{\tilde{x}_1=\tilde{x}_2^+}&=\psi_{nm}^-(\tilde{x}_1,\tilde{x}_2)|_{\tilde{x}_1=\tilde{x}_2^-};\\
\label{eq:NonLocalDelta2}\left(\frac{\partial}{\partial \tilde{x}_1}-\frac{\partial}{\partial \tilde{x}_2}-2\alpha\right)\psi_{mn}^+(\tilde{x}_1,\tilde{x}_2)|_{\tilde{x}_1=\tilde{x}_2^+}&=\left(\frac{\partial}{\partial \tilde{x}_1}-\frac{\partial}{\partial \tilde{x}_2}\right)\psi_{nm}^-(\tilde{x}_1,\tilde{x}_2)|_{\tilde{x}_1=\tilde{x}_2^-}
\end{align}
for all $\tilde{x}_1,\tilde{x}_2,\in(0,l_{mn}^-)$ when $(i_m,i_n)\in\mathcal{N}$.

As in the equilateral star case, using the method devised in \cite{BKContact} one can introduce vectors of boundary values that reproduce the boundary conditions \eqref{eq:NonLocalDelta1}--\eqref{eq:NonLocalDelta2} and prove that they provide a self-adjoint realisation of the two-particle Laplacian.

\subsection{Spectra}\label{sec:GeneralSpectra}

We have seen how to establish boundary conditions which correspond to two-particle quantum graphs with $\tilde{\delta}$-type interactions. We would now like to show that such systems are exactly solvable and calculate their spectra.

For the equilateral star graphs considered in Section \ref{sec:2PES}, exact solvability was shown by substituting the ansatz \eqref{eq:EquiAnsatz} directly into boundary conditions \eqref{eq:DeltaV} and \eqref{eq:TildeDelta1}--\eqref{eq:TildeDelta2} defined on $D_{\Gamma_e}^*$. The spectra then followed by generalising the approach in \cite{KosSch05} to two particles. While, in principle, we can use the same method in the general graph case, the extra complexity brought about by different edge lengths makes the presentation rather convoluted. To this end we will use the method presented in Section \ref{sec:1pScatteringMatrices} which utilises the star representation $\Gamma^{(s)}$ of a compact graph $\Gamma$. In the compact setting, $\tilde\delta$-interactions require us to define dissections of the domains $D_{mn}$ when they come from neighbouring pairs $(i_m,i_n)$ of edges on $\Gamma$. On $\Gamma^{(s)}$, this corresponds to defining dissections of
\begin{align}
{D}_{mn}^{(s)}=(0,\infty)\times (0,\infty),
\end{align}
with $f(e_m)=f(e_n)$, according to
\begin{align}\label{eq:nmdissecstar+}
{D}_{mn}^{(s,+)}=\{(x_1,x_2)\in {D}^{(s)}_{mn};x_1>x_2\},
\end{align}
and
\begin{align}\label{eq:nmdissecstar-}
{D}_{mn}^{(s,-)}=\{(x_1,x_2)\in {D}_{mn}^{(s)};x_1<x_2\}.
\end{align}
This yields the dissected configuration space
\begin{align}\label{eq:GeneralSRepDissection}
D_{\Gamma^{(s)}}^*=\left(\bigcup_{m,n\atop f(e_m)\neq f(e_n)}{D}_{mn}^{(s)}\right)\bigcup
\left(\bigcup_{m,n\atop f(e_m)=f(e_n)}\left({D}_{mn}^{(s,+)}\cup {D}_{mn}^{(s,-)}\right)\right),
\end{align}
and the Hilbert space
\begin{align}
\mathcal{H}_2^{(s,*)}=L^2(D_{\Gamma^{(s)}}^*).
\end{align}
The task is now to specify the boundary conditions we would like to impose. Interactions at the vertices in this setting will be described by simple two-particle lifts of \eqref{eq:LocalBoundaryConditions} and are exact the analogues of \eqref{eq:ContactVVector}--\eqref{eq:DeltaV}. The $\tilde{\delta}$-interactions are implemented through
the conditions
\begin{align}\label{eq:StarRepP1}
\psi_{mn}^{(s,+)}(x_1,x_2)|_{x_1=x_2^+}&=\psi_{nm}^{(s,-)}(x_1,x_2)|_{x_1=x_2^-};\\
\label{eq:StarRepP2}
\left(\frac{\partial}{\partial x_1}-\frac{\partial}{\partial x_2}-2\alpha\right)\psi_{mn}^{(s,+)}(x_1,x_2)|_{x_1=x_2^+}&=\left(\frac{\partial}{\partial x_1}-\frac{\partial}{\partial x_2}\right)\psi_{nm}^{(s,-)}(x_1,x_2)|_{x_1=x_2^-}.
\end{align}
on edge pairs $(e_m,e_n)$ where $f(e_m)=f(e_n)$. 

We note that below an alternative notation will be more convenient. We shall artificially extend the dissections \eqref{eq:nmdissecstar+}--\eqref{eq:nmdissecstar-} to all edge pairs and correct for this by setting 
\begin{align}\label{eq:StarRepNonInt1}
\psi_{mn}^{(s,+)}(x_1,x_2)|_{x_1=x_2^+}&=\psi_{mn}^{(s,-)}(x_1,x_2)|_{x_1=x_2^-};\\
\label{eq:StarRepNonInt2}
\left(\frac{\partial}{\partial x_1}-\frac{\partial}{\partial x_2}\right)\psi_{mn}^{(s,+)}(x_1,x_2)|_{x_1=x_2^+}&=\left(\frac{\partial}{\partial x_1}-\frac{\partial}{\partial x_2}\right)\psi_{mn}^{(s,-)}(x_1,x_2)|_{x_1=x_2^-}.
\end{align}
for edge pairs with  $f(e_m)\neq f(e_n)$. Using the Bethe ansatz method, an eigenvector $\Psi^{(s)}$ in the star representation will then be described by the collection of functions
\begin{align}
\psi_{mn}^{(s,\pm)}(x_1,x_2)=\sum_{P\in\mathcal{W}_2}\mathcal{A}_{mn}^{(P,\pm)} e^{i(k_{P1} x_1+k_{P2} x_2)}
\end{align}
on $D_{mn}^{(s,\pm)}$ (for all edge pairs). Let us define the $2|\mathcal{E}|^2$-dimensional vector
\begin{align}
\mathcal{A}^P=\begin{pmatrix}
(\mathcal{A}_{mn}^{(P,-)})_{m,n=1}^{|\mathcal{E}|}\\
\mathbb{T}_{|\mathcal{E}|^2}(\mathcal{A}_{mn}^{(PT,+)})_{m,n=1}^{|\mathcal{E}|}
\end{pmatrix},
\end{align}
with $\mathbb{T}$ as defined in \eqref{eq:Tmatrixdef}. The vertex-boundary conditions then imply
\begin{align}\label{eq:GeneralPR}
\mathcal{A}^{PR}=\left(\mathbb{I}_2\otimes S_v(-k_{P1})\otimes\mathbb{I}_{|\mathcal{E}|}\right)\mathcal{A}^P
\end{align}
for all $P\in\mathcal{W}_2$. At this point, it is convenient to define the matrix $\boldsymbol{c}=\diag (c_{mn})_{{m,n=1}}^{|\mathcal{E}|}$ where 
\begin{align}\label{eq:cGeneral}
c_{mn}=\begin{cases} 1 &\text{if } f(e_m)=f(e_n);\\ 
0 &\text{else}, 
\end{cases}
\end{align}
which distinguishes domains with $\tilde{\delta}$-type interactions from those where derivatives are continuous across dissections. The $\tilde\delta$-conditions then imply
\begin{align}\label{eq:GeneralPT}
\mathcal{A}^{PT}=Y(k_{P1}-k_{P2})\mathcal{A}^P
\end{align}
with
\begin{align}\label{eq:GeneralY}
Y(k)=S_p(k)\otimes \boldsymbol{c}+
\begin{psmallmatrix}
0&1\\
1&0
\end{psmallmatrix}\otimes
(\mathbb{I}_{|\mathcal{E}|^2}-\boldsymbol{c})\mathbb{T}_{|\mathcal{E}|^2}.
\end{align}

To prove exact solvability we need only show that relations \eqref{eq:GeneralPR} and \eqref{eq:GeneralPT} are consistent with the properties of $\mathcal{W}_2$. This amounts to the requirements
\begin{enumerate}
\item $S_v(u)S_v(-u)=\mathbb{I}_{|\mathcal{E}|}$;
\item $Y(k)Y(-k)=\mathbb{I}_{2|\mathcal{E}|}$;
\item $\left(\mathbb{I}_2\otimes S_v(u)\otimes \mathbb{I}_{|\mathcal{E}|} \right)Y(u+v)\left(\mathbb{I}_2\otimes S_v(v)\otimes\mathbb{I}_{|\mathcal{E}|}\right)Y(v-u)\\
=Y(v-u)\left(\mathbb{I}_2\otimes S_v(v)\otimes \mathbb{I}_{|\mathcal{E}|} \right)Y(u+v)\left(\mathbb{I}_2\otimes S_v(u)\otimes\mathbb{I}_{|\mathcal{E}|}\right)$.
\end{enumerate}
The first two conditions are easily verified by the explicit forms of $S_v(u)$ and $Y(u)$, noting that, since $c_{mn}=c_{nm}$, the properties of $\mathbb{T}_{|\mathcal{E}|^2}$ are such that
\begin{align}
[\boldsymbol{c},\mathbb{T}_{|\mathcal{E}|^2}]=0.
\end{align}
Noting then the relation
\begin{align}
[S_v(u)\otimes \mathbb{I}_{|\mathcal{E}|},\boldsymbol{c}]=0
\end{align}
holds if vertex boundary conditions are local, that is $S_v(u)$ obeys the condition \eqref{eq:LocalAB}, and also the relation \eqref{eq:EquiKSCommutation}, the third condition is easily verified.

Let us bring our attention back to the original compact graph $\Gamma$. In order to turn the eigenfunctions in the star representation into eigenfunctions on the compact graph, it is sufficient to impose the relations
\begin{align}
&\psi^{(s,+)}_{mn}(x_1,x_2)=\psi_{(m+|\mathcal{I}|)n}^{(s,+)}(l_m-x_1,x_2) \quad\text{and}\quad
&\psi^{(s,-)}_{mn}(x_1,x_2)=\psi_{m(n+|\mathcal{I}|)}^{(s,-)}(x_1,l_n-x_2)
\end{align}
for all $m,n\in\{1,..,|\mathcal{I}|\}$ which imply
\begin{align}
&\mathcal{A}_{mn}^{(P,+)}=\mathcal{A}_{(m+|\mathcal{I}|)n}^{(PR,+)}e^{-ik_{P1}l_m}\quad\text{and}\quad
&\mathcal{A}_{mn}^{(P,-)}=\mathcal{A}_{m(n+|\mathcal{I}|)}^{(PTRT,-)}e^{-ik_{P2}l_n}.
\end{align}
These conditions then yield the relation
\begin{align}\label{eq:BasisRelation}
\mathcal{A}^{P}=E(-k_{P2})
\mathcal{A}^{PTRT}
\end{align}
where
\begin{align}
E(k)=\mathbb{I}_{4|\mathcal{I}|}\otimes\begin{pmatrix}
0&1\\
1&0
\end{pmatrix}
\otimes
e^{ik\boldsymbol{l}}
\end{align}
with $e^{ik\boldsymbol{l}}$ defined as in \eqref{eq:Boldl}. Applying \eqref{eq:GeneralPR}, \eqref{eq:GeneralPT} and \eqref{eq:BasisRelation} successively we have the condition that
\begin{align}\label{eq:GenealSecEqu}
Z(k_{P1},k_{P2})=0
\end{align}
with
\begin{align}\label{eq:FinalEightQCon}
Z(k_1,k_2)=\det\left[\mathbb{I}_{8|\mathcal{I}|^2}-E(k_{2})Y(k_2-k_1)\left(\mathbb{I}_2\otimes S_v(k_2)\otimes\mathbb{I}_{2|\mathcal{I}|}\right)Y(k_{1}+k_{2})\right]
\end{align}
is satisfied for all $P\in\mathcal{W}_2$. By using properties of determinants, the commutation relations established above and the explicit forms of $Y(k)$, $S_v(k)$ and $E(k)$, it is easy to see that the form \eqref{eq:FinalEightQCon} is such that if \eqref{eq:GenealSecEqu} is satisfied for some $P\in\mathcal{W}_2$, then it is necessarily satisfied for elements $PR,PTRT\in\mathcal{W}_2$. With this in mind we can state the main result of this section.
\begin{thm}\label{thm:GeneralMainthm}
Eigenvalues of a self-adjoint two-particle Laplacian $-\Delta_2$ defined on $\Gamma$ with local vertex interactions specified through $A,B$ and $\tilde{\delta}$-type interactions between particles when they are located on neighbouring edges are the values $E=k_1^2+k_2^2\neq 0$ with multiplicity $m$ where $k_1,k_2$ are simultaneous solutions to the secular equations
\begin{align}\label{eq:FinalQCon}
Z(k_i,k_j)=0
\end{align}
for $j,i\neq j\in\{1,2\}$ with multiplicity $m$.
\end{thm}

\subsection{Recovering specific results}\label{sec:specresult}

To finish the section, we establish agreement between the spectra of general two-particle quantum graphs presented in Theorem \ref{thm:GeneralMainthm} and results derived and discussed earlier in the paper.

The above result matches our previous result for equilateral stars given by Theorem \ref{thm:EquiSpectrum} by choosing the same boundary conditions at the vertices and setting all edge lengths to $l_m=l$. As, on star graphs, $\tilde\delta$-interactions are defined between all pairs of edges, one has to put
\begin{align}\label{eq:Equice}
c_{mn}=\begin{cases} 1 &\text{if } f(e_m)=f(e_n)=v_1 \text{ or } f(e_m),f(e_n)\in\{v_2,\dots,v_{|\mathcal{V}|}\};\\ 
0 &\text{else}.
\end{cases}
\end{align}
Substituting these parameters into $Z(k_i,k_j)$ and using the properties of determinants we recover the form $Z_e(k_i,k_j)$ as required.

In Section \ref{sec:2PES}, we introduced the notion of pseudo-non-interacting particles and showed that in the equilateral star setting, the appropriate quantisation condition is indeed that of the truly non-interacting case. However, in the general setting, this agreement does not hold. The spectra of such systems is calculated first by identifying the matrix
\begin{align}
\lim_{\alpha\rightarrow 0}Y(k)=
\begin{psmallmatrix}
0&1\\
1&0
\end{psmallmatrix}
\otimes
(\boldsymbol{c}+(\mathbb{I}_{|\mathcal{E}|^2}-\boldsymbol{c})\mathbb{T}_{|\mathcal{E}|^2}).
\end{align}
Substitution into \eqref{eq:FinalEightQCon} then yields the quantisation condition
\begin{align}\label{eq:GeneralPsuedoNonInt}
\begin{split}
Z(k)=\det\Big[\mathbb{I}_{4|\mathcal{I}|^2}-&\left(S_v(k)\otimes
\begin{psmallmatrix}
0&1\\
1&0 
\end{psmallmatrix}
\otimes e^{ik\boldsymbol{l}}\right)\boldsymbol{c}\\
-&\left(\mathbb{I}_{2|\mathcal{I}|}\otimes(
\begin{psmallmatrix}
0&1\\
1&0 
\end{psmallmatrix}
\otimes e^{ik\boldsymbol{l}})S_v(k)\right)\left(\mathbb{I}_{4|\mathcal{I}|^2}-\boldsymbol{c}\right)
\Big]
\end{split}
\end{align}
which we notice is dependent on the single momentum $k$.

Truly non-interacting systems are recovered by turning off all coupling between domains $D_{mn}$ and $D_{nm}$. This is achieved by setting $\boldsymbol{c}=\boldsymbol{0}$. We then have that
\begin{align}
Y(k)|_{\boldsymbol{c}=\boldsymbol{0}}=
\begin{psmallmatrix}
0&1\\
1&0
\end{psmallmatrix}
\otimes
\mathbb{T}_{|\mathcal{E}|^2}
\end{align}
By substituting into \eqref{eq:FinalEightQCon} we recover the secular equation \eqref{eq:1PSecular} for the one-particle quantum graph. 

In the subsequent section we would like to analyse examples of bosons on graphs. Computationally speaking, such examples are useful as the dimension of the matrix inside the determinant $Z(k_1,k_2)$ is halved. Imposing bosonic symmetry we have
\begin{align}
\psi_{mn}^{(s,\pm)}(x_1,x_2)=\psi_{nm}^{(s,\mp)}(x_2,x_1)
\end{align}
for all $(x_1,x_2)\in D_{mn}^{(s)}$ which implies the relations
\begin{align}
\mathcal{A}^{(P,-)}=\mathcal{A}^{(PT,+)}
\end{align}
for all $P\in\mathcal{W}_2$. The matrix \eqref{eq:GeneralY} then reduces to the form
\begin{align}
Y_b(k)=\mathbb{I}_2\otimes
\left(\frac{k-i\alpha}{k+i\alpha}\boldsymbol{c}+(\mathbb{I}_{|\mathcal{E}|^2}-\boldsymbol{c})\mathbb{T}_{|\mathcal{E}|^2}\right).
\end{align}
so that from \eqref{eq:FinalEightQCon}, we recover
\begin{align}\label{eq:SecB}
Z_b(k_1,k_2)=\det\left[\mathbb{I}_{4|\mathcal{I}|^2}-E_b(k_{2})Y_b(k_2-k_1)\left(S_v(k_2)\otimes\mathbb{I}_{2|\mathcal{I}|}\right)Y_b(k_{1}+k_{2})\right]
\end{align}
with $E(k)=\mathbb{I}_2\otimes E_b(k)$.

\section{Spectral Statistics}\label{sec:SS}

In this section we calculate the spectra of certain examples of two-particle quantum graphs and analyse their statistics. 

Let us first introduce the eigenvalue counting function
\begin{align}\label{eq:Count}
N(E)=\#\{n;\ E_n\leq E\}.
\end{align}
It was shown in \cite{BKContact} that for two-particle quantum graphs with $\delta$-type interactions, the asymptotic behaviour follows the Weyl law
\begin{align}\label{eq:WeylContact}
N(E)\sim\frac{\mathcal{L}^2}{4\pi}E,\ E\rightarrow \infty,
\end{align}
where $\mathcal{L}=\sum_{j=1}^{|\mathcal{I}|}l_j$ denotes the total length of the graph. Of course, the majority of this paper is not concerned with $\delta$-type interactions, but with $\tilde{\delta}$-interactions. Nonetheless, it is still revealing to compare the eigenvalue count with \eqref{eq:WeylContact}. To this end, we will assign a line of best fit
\begin{align}\label{eq:Fit}
\overline{N}(E)=aE+b\sqrt{E}+c
\end{align}
to the counting function and compare the leading term to \eqref{eq:WeylContact}.

One of the main motivations for the study of quantum graphs is to analyse their spectral correlations. A particularly useful statistical measure is the nearest neighbour level spacings distribution
\begin{align}\label{eq:Intp}
\int_a^b p(s) ds=\lim_{N\rightarrow \infty}\frac{1}{N}\#\{n\leq N;\ a\leq \epsilon_{n+1}-\epsilon_n\leq b\}.
\end{align}
of the unfolded versions, $\epsilon_1<\epsilon_2<\epsilon_3<\dots$, of the energy eigenvalues; that is the energies are rescaled such that the average spacing is equal to unity. Generic quantum systems with integrable classical limits are conjectured to have spectra with Poissonian statistics \cite{BT77},
\begin{align}
p(s)=e^{-s},
\end{align}
while chaotic classical systems have quantum counterparts with correlations described by random matrix models. For systems with integer spin and time-reversal symmetry Gaussian orthogonal ensemble (GOE) statistics are conjectured to apply \cite{BohGiaSch84},
where the level spacings distribution can be approximated by
\begin{align}
p(s) = \frac{\pi}{2}s\, \mathrm{e}^{-\frac{\pi}{4} s^2},
\end{align}
see \cite{Haa91}. In this section we analyse the spectral statistics of certain examples by using the integrated measure \eqref{eq:Intp}. 

In \cite{KS97}, nearest neighbour energy level distributions of one-particle quantum tetrahedra were shown to exhibit GOE spectral statistics and thus imply chaotic classical counterparts. In this section we analyse the spectra of two-particle quantum graphs calculated in the previous sections, looking for a potential dependence of spectral correlations on the interaction strength. We refer to the result by Srivastava et al. \cite{Sri16} who analysed the spectral properties of interacting kicked rotors which individually show GOE statistics. They found a transition from Poissonian to GOE statistics as the strength of the interaction was increased. 

\subsection{Tetrahedron}

Let us take, as an example, a system of two $\tilde{\delta}$-interacting bosons on a tetrahedron. The appropriate spectra are calculated according to Theorem \ref{thm:GeneralMainthm} using the secular equation \eqref{eq:SecB}. Vertex boundary conditions are determined by choosing Discrete Fourier Transform (DFT) scattering matrices $S_v^{(\eta,DFT)}$ with elements
\begin{align}\label{eq:DFT}
(S_v^{(\eta,DFT)})_{\gamma\lambda}=\frac{1}{\sqrt{d_\eta}}e^{2\pi i\frac{n(\gamma)n(\lambda)}{d_\eta}},
\end{align}
where $n(\cdot)$ is a bijection of the $d_\eta$ neighbouring  vertices of $v_\eta$ onto the numbers $\{0,\dots,d_\eta-1\}$. For the tetrahedron, appropriate DFT scattering matrices at each vertex $v_\eta$ are then
\begin{align}
S_\eta^{(DFT)}=\frac{1}{\sqrt{3}}
\begin{pmatrix}
1&1&1\\
1&e^{\frac{2i\pi}{3}}&e^{\frac{4i\pi}{3}}\\
1&e^{\frac{4i\pi}{3}}&e^{\frac{8i\pi}{3}}
\end{pmatrix},
\end{align}
with distinct eigenvalues $\{-1,1,i\}$. With this choice, the spectrum of the two-particle Laplacian with $\tilde\delta$-interactions is non-degenerate. 

\begin{figure}
\centering
\includegraphics{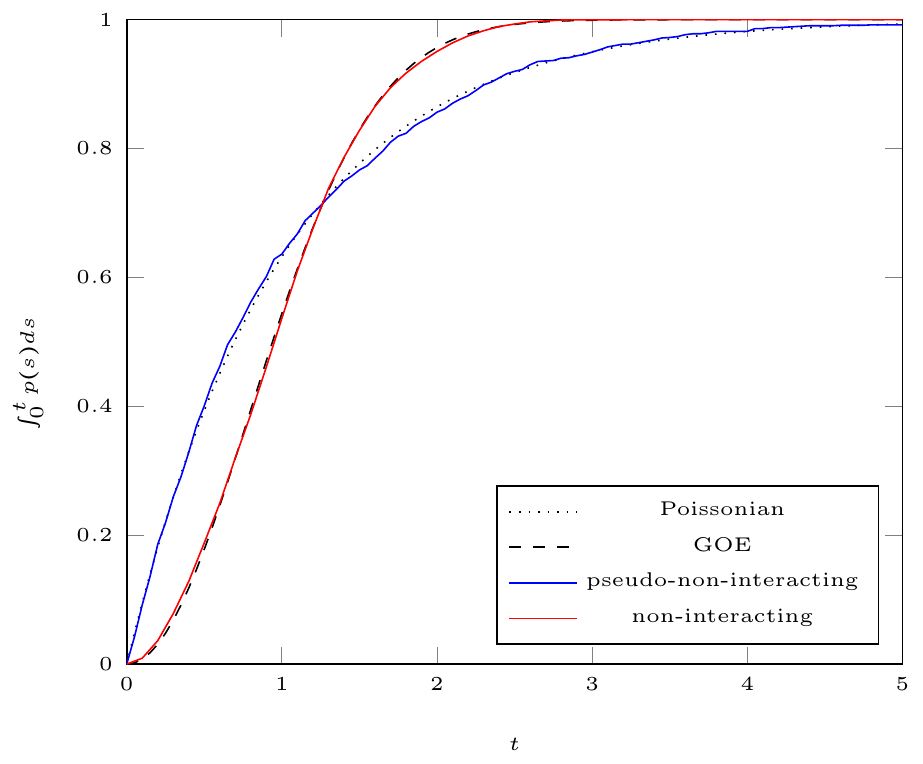}
\caption{Integrated level spacings distributions for systems of non-interacting and pseudo-non-interacting particles on a tetrahedron. First 50,000 eigenvalues.}
\label{fig:TetNI}
\end{figure}

\begin{figure}
\centering
\begin{minipage}[b]{.48\textwidth}
\includegraphics{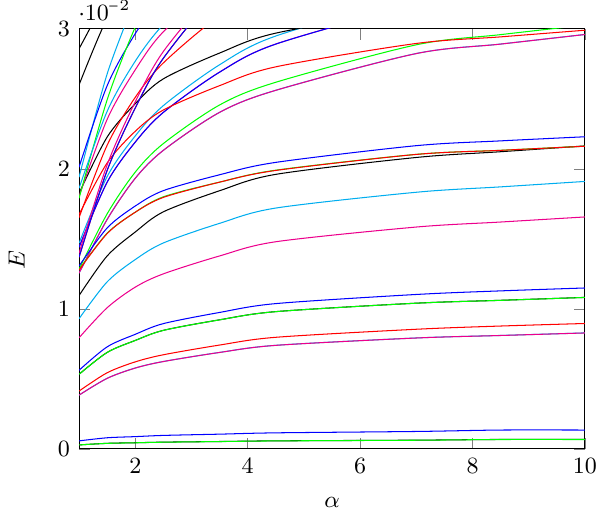}
\caption{Dependency on interaction strength of small eigenvalues of a system of two bosons on a tetrahedron.}
\label{fig:TetLevels}

\end{minipage}
\hspace{4pt}
\begin{minipage}[b]{0.48\textwidth}
\includegraphics{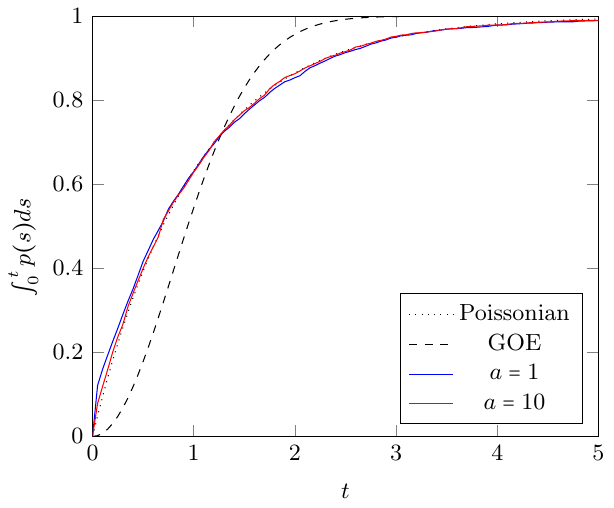}
\caption{Integrated level spacings distributions for systems of two bosons on a tetrahedron. First 3000 eigenvalues.}
\label{fig:TetNN}
\end{minipage}
\end{figure}

\begin{figure}
\centering
\includegraphics{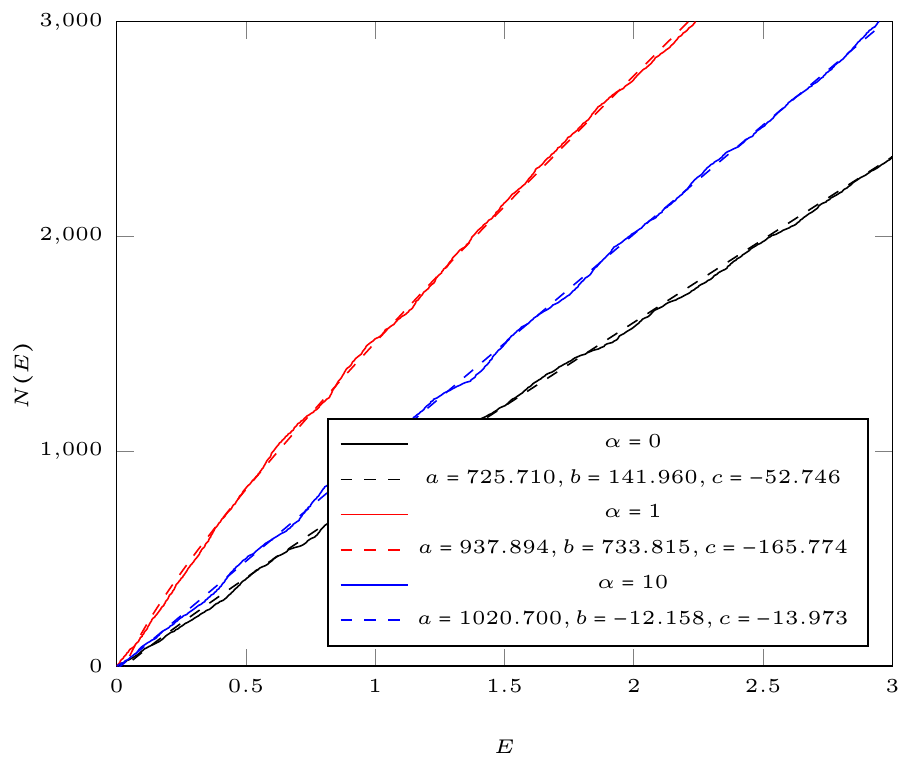}
\caption{Counting functions $N(E)$ (solid line) with lines of best fit $\overline{N}(E)$ (dashed line) for systems of two bosons on a tetrahedron.}
\label{fig:CountingTet}
\end{figure}

Before analysing two-particle spectra, let us consider the spectra of non-interacting systems. Figure \ref{fig:TetNI} plots the nearest neighbour distributions for the single-particle spectra associated with truly non-interacting ($\boldsymbol{c}=\boldsymbol{0}$) and pseudo-non-interacting ($\alpha=0$) particles on the tetrahedron with DFT scattering matrices. As is well-known \cite{KS97} and confirmed in Figure \ref{fig:TetNI}, the one-particle spectrum follows GOE statistics. The pseudo-non-interacting system, however, shows Poissonian statistics. The crucial point here is that two-particle systems prescribed in Theorem \ref{thm:GeneralMainthm} in fact couple systems of pseudo-non-interacting particles which individually possess spectra with Poissonian statistics, not systems of truly non-interacting particles which individually follow GOE statistics. Thus we cannot expect a transition to GOE statistics as in \cite{Sri16}. Figure \ref{fig:TetLevels} plots the $\alpha$-dependency of the lowest energy levels of a system of $\tilde{\delta}$-interacting bosons on a tetrahedron with DFT vertex scattering matrices. There is no obvious transition to a regime of energy level repulsion as we increase $\alpha$. Indeed, plots of  nearest neighbour distributions reveal Poissonian statistics for all interaction strengths. Figure \ref{fig:TetNN} shows these plots for interaction strengths $\alpha=1$ and $\alpha=10$. Figure \ref{fig:CountingTet} plots counting functions $N(E)$ for strengths $\alpha\in\{0,1,10\}$ together with quadratic lines of best fit \eqref{eq:Fit}. In each case, the leading term does not agree with the Weyl law \eqref{eq:WeylContact} predicted for contact interactions.

\subsection{Equilateral star}
\begin{figure}
\centering
\includegraphics{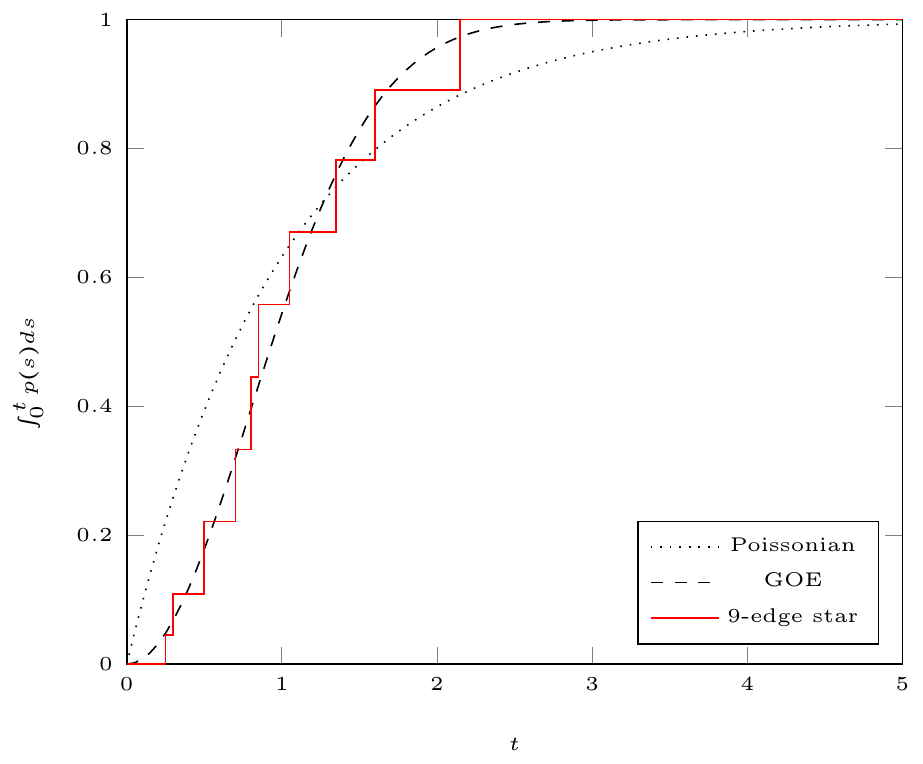}
\caption{Integrated level spacings distributions for systems of non-interacting and pseudo-non-interacting particles on a $9$-edge equilateral star. First 50,000 eigenvalues.}
\label{fig:StarNI}
\end{figure}

\begin{figure}
\centering
\begin{minipage}[b]{.48\textwidth}
\vspace{4pt}
\includegraphics{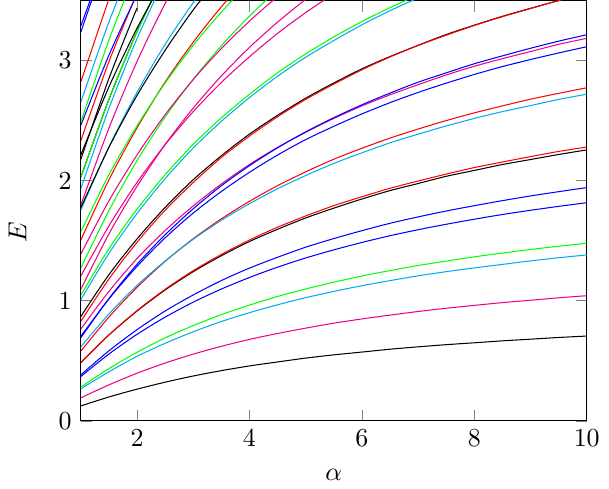}
\caption{Dependency on interaction strength of small eigenvalues of a system of two bosons on a $9$-edge equilateral star.}
\label{fig:StarLevels}
\end{minipage}
\hspace{4pt}
\begin{minipage}[b]{0.48\textwidth}
\includegraphics{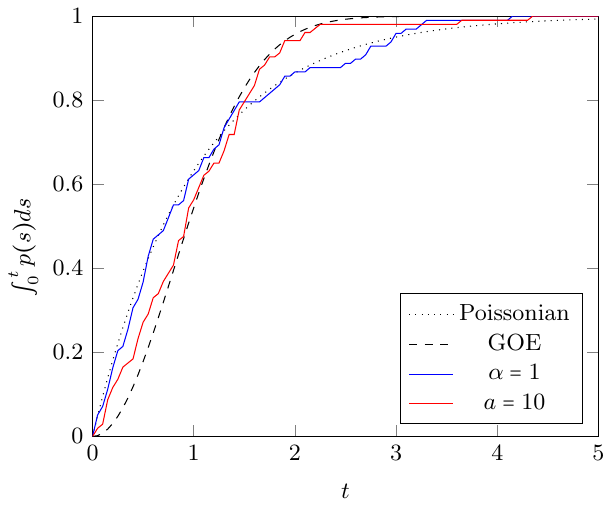}
\caption{Integrated level spacings distributions for systems of two bosons on a $9$-edge equilateral star. First $100$ eigenvalues.}
\label{fig:StarNN}
\end{minipage}
\end{figure}

\begin{figure}
\centering
\begin{minipage}[b]{.48\textwidth}
\includegraphics{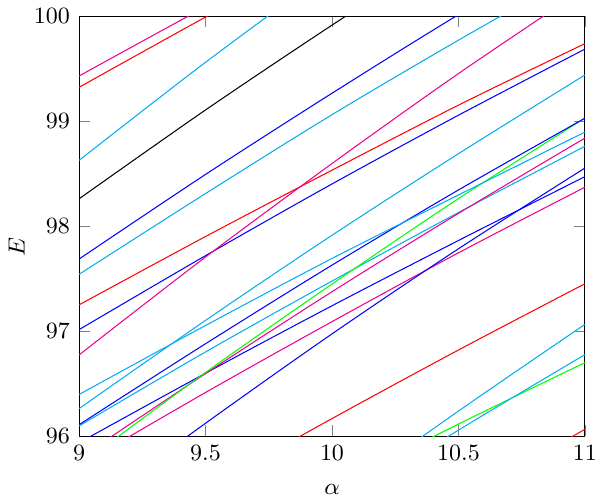}
\caption{Dependency on interaction strength of large eigenvalues of a system of two bosons on a $9$-edge equilateral star.}
\label{fig:StarLevelsMore}
\end{minipage}
\hspace{4pt}
\begin{minipage}[b]{0.48\textwidth}
\includegraphics{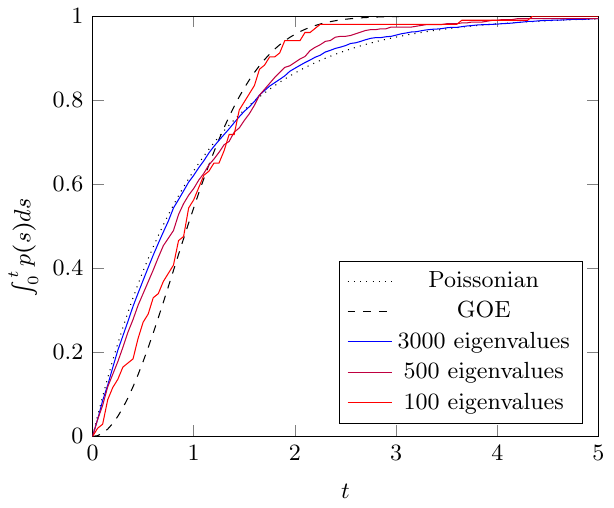}
\caption{Integrated level spacings distributions for systems of two bosons on a $9$-edge equilateral star with $\alpha=10$.}
  \label{fig:StarNNMore}
\end{minipage}
\end{figure}  

To examine the spectral statistics of coupled chaotic systems we must look for two-particle quantum graphs for which the one-particle spectra recovered when setting $\alpha=0$ are chaotic. We have seen that the two-particle tetrahedron does not fulfil this requirement. Let us then focus our attention on equilateral stars which we discussed in Section \ref{sec:2PES}, where we showed that we recover true one-particle spectra when setting $\alpha=0$. Thus we can discuss coupled chaotic systems in the spirit of \cite{Sri16} if we can find one-particle equilateral stars which exhibit GOE statistics. Such systems are characterised by the quantisation condition \eqref{eq:EquiQCForma0} which can be written
\begin{align}
e^{-2ikl}=-\mu(k),
\end{align}
where $\mu(k)$ is an eigenvalue of $S^{(1)}_v(k)$. Clearly, the multiplicity of solutions $k$ are equal to the multiplicity these eigenvalues. For example, equilateral stars with boundary conditions characterised by the DFT scattering matrix \eqref{eq:DFT} at the central vertex yield solutions corresponding to $\mu=\{1,-1,i,-i\}$ with degenerate values arising for $d>3$. Clearly degenerate energy levels would obscure conclusions made in the context of spectral statistics. To navigate this issue, we must choose a scattering matrix with non-degenerate eigenvalues. In what follows, we choose a $d_{|\mathcal{I}|}\times d_{|\mathcal{I}|}$ random unitary matrix. Figure \ref{fig:StarNI} plots the nearest neighbour distribution for a single particle on such an equilateral star with $9$ edges. The degenerate energy level spacings arise from the imposition of equal lengths. Indeed in studies of one-particle quantum graph spectra, rationally independent lengths are chosen to avoid degenerate level spacings. We do however see approximate agreement with GOE statistics. In this setting we can thus investigate the coupling of two chaotic spectra by increasing $\alpha$ from $0$.

\begin{figure}
\centering
\includegraphics{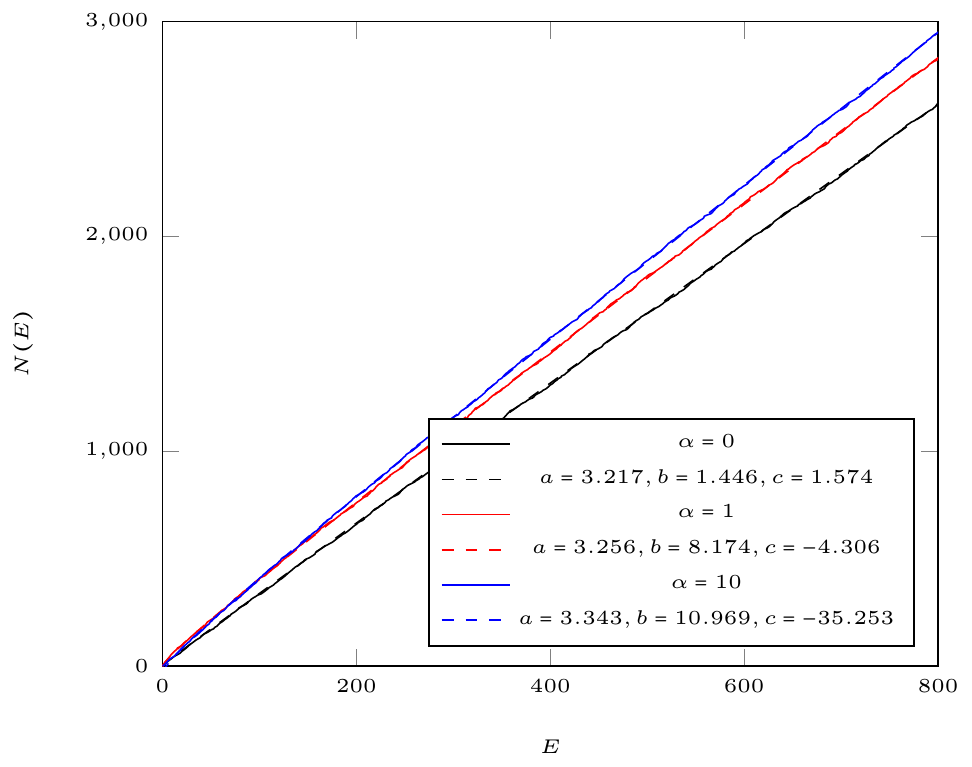}
\caption{Counting functions $N(E)$ (solid line) with lines of best fit $\overline{N}(E)$ (dashed line) for systems of two bosons on a $9$-edge equilateral star.}
\label{fig:CountingStar}
\end{figure}

Figure \ref{fig:StarLevels} plots the $\alpha$-dependency of the lowest energy levels of a system of two bosons on a $9$-edge equilateral star with a random unitary central scattering matrix. We clearly see a transition to level repulsion as $\alpha$ is increased. Figure \ref{fig:StarNN} plots nearest neighbour distributions for the first $100$ energy levels. There is a clear shift from Poissonian, for $\alpha=1$, to GOE statistics, for $\alpha=10$. We note, however, that this level repulsion becomes less apparent as we include larger energy levels; Figure \ref{fig:StarLevelsMore} shows level crossing at higher energies and Figure \ref{fig:StarNNMore} shows how the spectral statistics for the $\alpha=10$ case tend to Poissonian as we include higher energies. Figure \ref{fig:CountingStar} plots counting functions for $\alpha\in\{0,1,10\}$ together with quadratic lines of best fit \eqref{eq:Fit}. For the non-interacting ($\alpha=0$) case, the leading term agrees exactly with the Weyl law \eqref{eq:WeylContact}. As the interaction strength increases, the counting function diverges from this prediction.

\small{\bibliographystyle{amsalpha}
\bibliography{Literature}}

\end{document}